\documentclass[twocolumn,aps,showpacs,superscriptaddress,prl]{revtex4}
\pdfoutput=1
\usepackage{ifpdf}
\usepackage{graphicx}
\usepackage{times}
\usepackage{amsmath}
\usepackage{amssymb}
\usepackage{units}
\usepackage{stmaryrd} 

\renewcommand{\vec}[1]{\boldsymbol{#1}}

\begin{document}
\preprint{0}

\title{Tunable spin-gaps in a quantum-confined geometry}

\author{Emmanouil Frantzeskakis}
\affiliation{Laboratoire de Spectroscopie \'Electronique, Institut de
  Physique des Nanostructures, \'Ecole Polytechnique F{\'e}d{\'e}rale
  de Lausanne (EPFL), station 3, CH-1015 Lausanne - Switzerland}

\author{St{\'e}phane Pons}
\email{stephane.pons@epfl.ch}
\affiliation{Laboratoire de
Spectroscopie \'Electronique, Institut de
  Physique des Nanostructures, \'Ecole Polytechnique F{\'e}d{\'e}rale
  de Lausanne (EPFL), station 3, CH-1015 Lausanne - Switzerland}
\affiliation{Laboratoire de Physique des Mat{\'e}riaux,
  Nancy-Universit{\'e}, CNRS, Boulevard des Aiguillettes, B.P. 239,
  F-54506 Vandoeuvre l{\`e}s Nancy, France}

\author{Hossein Mirhosseini}
\affiliation{Max-Planck-Institut f\"ur Mikrostrukturphysik,
Weinberg
  2, D-06120 Halle (Saale), Germany}

\author{J\"urgen Henk}
\affiliation{Max-Planck-Institut f\"ur Mikrostrukturphysik,
Weinberg
  2, D-06120 Halle (Saale), Germany}

\author{Christian. R. Ast}
\affiliation{Max-Planck-Institut f{\"u}r Festk{\"o}rperforschung, D-70569
Stuttgart, Germany}

\author{Marco Grioni}
\affiliation{Laboratoire de Spectroscopie \'Electronique, Institut de
  Physique des Nanostructures, \'Ecole Polytechnique F{\'e}d{\'e}rale
  de Lausanne (EPFL), station 3, CH-1015 Lausanne - Switzerland}


\begin{abstract}
We have studied the interplay of a giant spin-orbit splitting and
of quantum confinement in artificial Bi-Ag-Si trilayer structures.
Angle-resolved photoelectron spectroscopy (ARPES) reveals the
formation of a complex spin-dependent gap structure, which can be
tuned by varying the thickness of the Ag buffer layer. This
provides a means to tailor the electronic structure at the Fermi
energy, with potential applications for silicon-compatible
spintronic devices.
\end{abstract}

\maketitle In nonmagnetic centrosymmetric bulk solids like
silicon, electronic states of opposite spin have the same energy.
A surface or an interface breaks the translational invariance of a
three-dimensional crystal. Thus, as predicted by Bychkov and
Rashba\cite{Bychkov1984}, the spin-orbit (SO) interaction can lead
to spin-split electronic states in two-dimensional electron gases
(2DEG), in asymmetric quantum wells\cite{Koga2002}, at a surface
or at an interface\cite{Lashell1996,Rotenberg1999}. The size of
the splitting is related to the strength of the atomic SO coupling
(i. e. to the gradient of the atomic potential\cite{Malterre2007})
and to the potential gradient perpendicular to the
confinement\cite{Forster2004}. An unexpectedly large splitting was
recently reported for a BiAg surface alloy grown on a Ag(111)
single crystal\cite{Ast2007}. It is attributed to an additional
in-plane gradient of the surface potential, hence being a direct
consequence of the chemical alloy
configuration\cite{Ast2007,Premper2007}.

The spin-orbit interaction could be used to control via a gate
voltage the dynamics of spins injected into a
semiconductor\cite{Datta1990,Awshalom2007,Koga2002,Nitta1997}.
Moreover, the spin Hall effect - also induced by the SO
interaction - could find applications in new spintronic
devices\cite{Kato2004,Valenzuela2006} which rely neither on
magnetic materials nor on optical pumping. Interfaces between
silicon and materials exhibiting large spin-orbit splitting are
therefore expected to open novel vista for spintronics. The
challenge is to control the electronic states and spin
polarization at the Fermi level which determine the electron and
spin transport through interfaces\cite{Greullet2007,Tsymbal2007}
and nanostructures. Among the heavy metals which exhibit strong
spin-orbit interactions, bismuth may be favored for environmental
considerations. Experiments on thin layers of bismuth on silicon
have evidenced a SO splitting in the Bi surface states, but not of their bulk counterparts \cite{Hirahara2006,Hirahara2007}.
Moreover, it was observed that the
splitting  is removed by the hybridization between surface and bulk states.

\begin{figure}
  \centering
  \includegraphics[width = 8 cm]{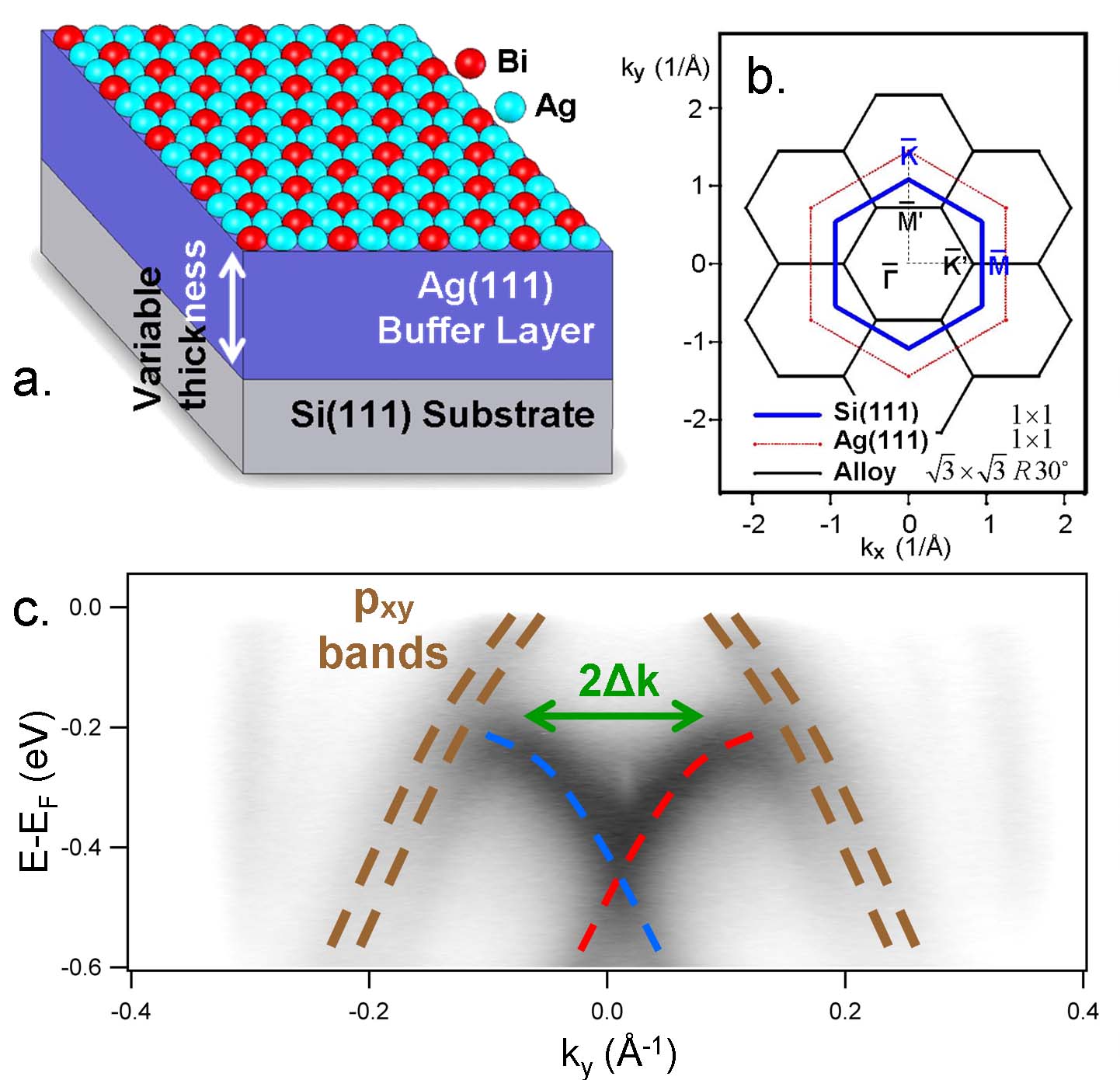}
  \caption{(color online) a, schematic view of a trilayer sample.
  The $\sqrt 3 \times \sqrt 3 R 30^{\circ}$ BiAg alloy is grown on a Ag buffer - whose thickness can be varied - deposited on a silicon substrate.
  b, first Brillouin zones of the surface structures.
  The symmetry lines $\bar \Gamma \bar K \bar M$ and $\bar \Gamma \bar K' \bar M'$ refer to Si(111) and to the alloy, respectively.
  c, ARPES intensity of the surface states of a Bi-Ag alloy grown on a thick Ag layer deposited on Si(111) along  . This system is similar to the alloy grown on a Ag(111) single crystal.
  Two parabolic bands of opposite spin with sp$_z$ character are split by $\Delta k = \pm 0.13${\AA}$^{-1}$ and cross at around -400meV at $\bar \Gamma$.
    Side bands of p$_x$p$_y$ symmetry pass through the Fermi level.
Close to $\bar \Gamma$ all bands exhibit a rotational symmetry
around the surface normal. The dashed lines are guides to the
eyes.}
  \label{fig1}
\end{figure}

In this Letter we explore a different approach. We
fabricated trilayer systems composed of a BiAg surface alloy\cite{Ast2007}, a
thin Ag buffer layer of variable
thickness (d), and a Si(111) substrate (Fig. 1a). Along the z
direction, the vacuum/BiAg/Ag/Si related potential is asymmetric
and SO splitting of delocalized electronic states is expected.
The good interfacial adhesion of the silver film makes
the system stable at room temperature (RT) and results in a sharp
interface.
We investigated the complex interface by
angle-resolved
photoelectron spectroscopy (ARPES) experiment, supported
by first-principles electronic-structure calculations.
We find that the SO splitting is large. We also find that, due to quantum confinement in the buffer layer, the
electronic structure exhibits patches of highly spin-polarized spectral
density. The spin-dependent density of states close to
the Fermi energy can be tuned by the thickness of the Ag buffer.

The experiments were performed with a multi-chamber set-up under
ultra-high vacuum. During preparation, Si(111) (highly phosphorus
doped, resistivity 0.009 - 0.011$\Omega.cm$) was flashed at
1200$^{\circ}$C by direct current injection. After the flashes,
the substrate was cooled slowly in order to obtain a sharp 7
$\times$ 7 signature in low-energy electron diffraction (LEED).
The Ag films were deposited with a home made Knudsen cell while
the sample was kept at 80K and then annealed at 400K. The quality
of the silver thin film was checked by LEED. Ag grows in the [111]
direction \cite{Speer2006}. The $\sqrt 3 \times \sqrt 3 R 30^{\circ}$ BiAg surface
alloy was obtained by depositing 1/3ML of Bi with an EFM3 Omicron
source on the sample at RT followed by a soft annealing.
Angle-resolved photoemission spectroscopy (ARPES) spectra were
acquired at RT and 55K with a PHOIBOS 150 Specs Analyser. We used
a monochromatized and partially polarized GammaData VUV 5000 high
brightness source of 21.2eV photons.

The first-principles
electronic-structure calculations are based on the local
spin-density approximation to density functional theory, as
implemented in relativistic multiple-scattering theory
(Korringa-Kohn-Rostoker and layer-Korringa-Kohn-Rostoker methods;
for details, see refs. \cite{Ast2007,Ast2008}). Spin-orbit coupling
is taken into account by solving the Dirac equation. The used
computer codes consider the boundary conditions present in
experiment, that is the semi-infinite substrate, a buffer of
finite thickness, the surface, and the semi-infinite vacuum. The
potentials of all sites (atoms) are computed self-consistently,
except for the Si substrate which is mimicked by spherical
repulsive potentials of 1 Hartree height. This so-called
hard-sphere substrate follows the face-centered cubic structure of
the Ag buffer. The electronic structure is addressed in terms of
the spectral density which is obtained from the imaginary part of
the Green function of the entire system. The latter can be
resolved with respect to wavevector, site, spin and angular
momentum, thus allowing a detailed analysis of the local
electronic structure.

The surface electronic properties of the alloy grown on top of a
thick Ag film (d=80 monolayers (ML)), as obtained by ARPES (Fig.
1c), agree with those of the alloy grown on a Ag(111) single
crystal\cite{Ast2007}. The spin-split bands which belong to
electronic states with sp$_z$ character cross at $\bar \Gamma$
(in-plane wavevector  $\vec {k_\sslash}=\vec 0$). They are well
described by parabolas (effective mass m* = -0.35 m$_e$) which are
offset by $\Delta k = \pm 0.13$\AA$^{-1}$. This shift in
wavevector is a signature of the aforementioned Rashba effect. The
set of side bands stems from electronic states of mainly
p$_x$p$_y$ character which are also spin-polarized but less
split\cite{Ast2007,Premper2007,Meier2008}. Electronic-structure
calculations\cite{Ast2007} show that the BiAg surface states are
much more strongly localized in the top layer than the Ag(111) or
Au(111) Shockley surface states. Thus, the spin-split bands and
the giant SO splitting are not directly affected by the Ag/Si
interface for Ag film thickness larger than a few monolayers. This
implies that prior results for BiAg/Ag(111)\cite{Ast2007} can be
transferred to silicon technology (i.e. to BiAg/Ag/Si(111)) at RT.
\begin{figure}
  \centering
  \includegraphics[width=8.6 cm]{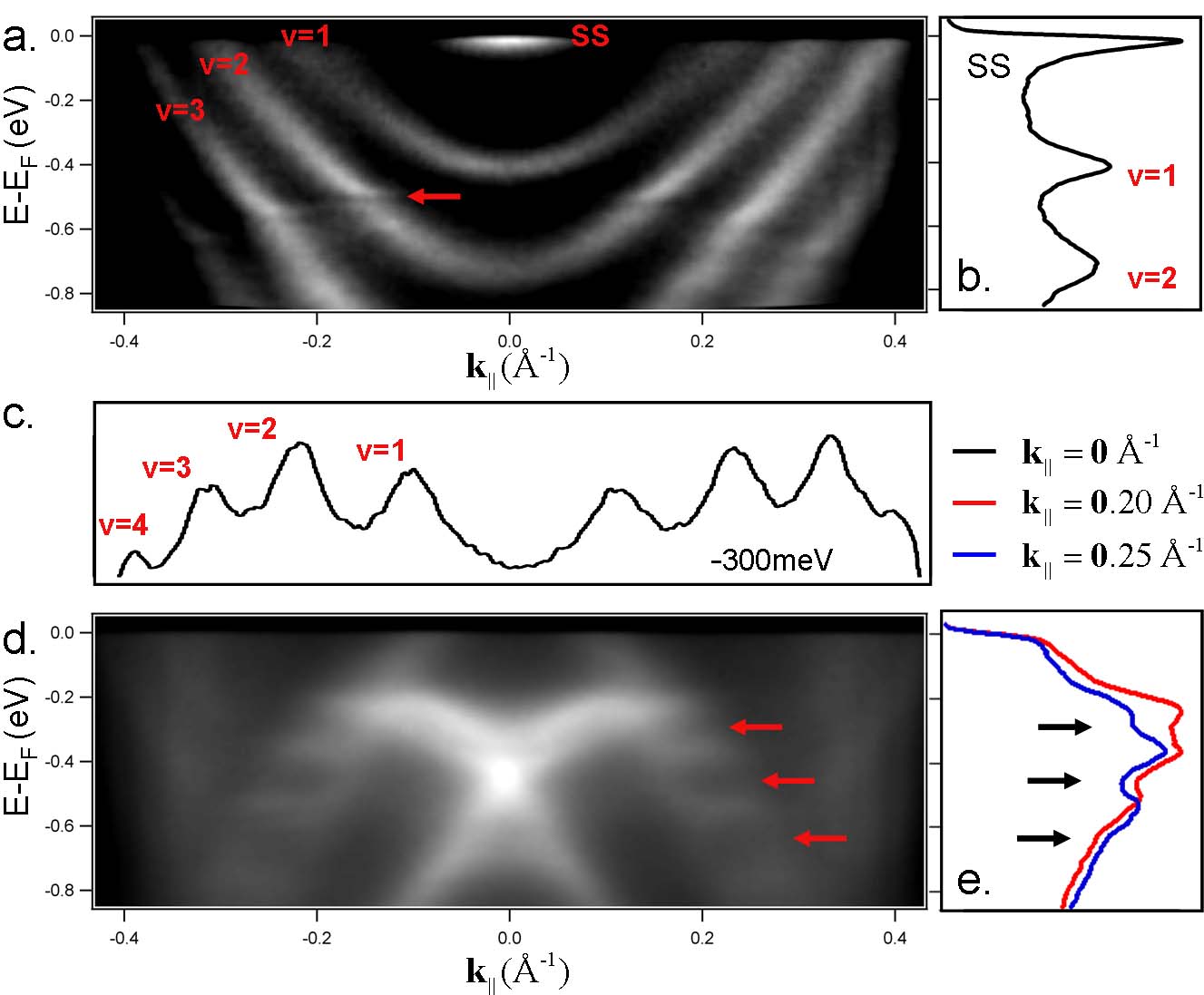}
  \caption{(color online) a, raw ARPES data along $\bar \Gamma \bar K$ at 55K. Quantum-well states originating from the confinement of the bulk sp bands in the 17ML bare Ag buffer deposited on Si(111).
  These states numbered~$\nu=1...n$~show up as a set of parabolic bands whose energy levels vary with the thickness of the film.
  Kinks in the dispersion (arrow) are due to the hybridization of the QWS with the p bands of silicon.
  SS stands for the Shockley surface states of Ag(111).
  b, EDC extracted from Fig. 2a at $\bar \Gamma$, i.e. k = 0.0{\AA}$^{-1}$. The 1$^{st}$ and 2$^{nd}$ QWS signatures and the SS are indicated.
  c, MDC extracted from Fig. 2a at -300meV shows the successive branches of the QWS.
  d, raw ARPES intensity along $\bar \Gamma \bar M'$ at 55K of the BiAg alloy grown on 17ML of Ag.
  e, EDC extracted from Fig. 2d for k = 0.20{\AA}$^{-1}$ and k = 0.25{\AA}$^{-1}$. Arrows indicate gaps of 100 - 200meV in the dispersion of the p$_x$p$_y$ bands.}
  \label{fig2}
\end{figure}

A new and interesting situation arises for thinner Ag buffer layers, where d is of the order of the
attenuation length of the electronic states. The Ag sp states are
confined to the Ag film by the potential barrier (image-potential
barrier) on the vacuum side (surface) and by the fundamental band
gap of Si on the substrate side. This confinement leads to
quantized wavevectors along z and to discrete energy
levels\cite{Chiang2000}. These so-called quantum well states
(QWS's) play a central role in transport
properties\cite{Jalochowski1992} and in the coupling of magnetic
layers in superlattices\cite{Ortega1992,Ortega1993,Bruno1995}.
Ag/Si(111) QWS's, in particular, have been extensively studied by
ARPES\cite{Speer2006,Wachs1986,Asensio2002}. For Ag(111) films, their in-plane
dispersion consists of a set of parabolic bands centered at $\bar
\Gamma$, with energies determined by the film thickness (Fig. 2a;
d=17 ML). The electronic fringe structure with a negative
parabolic dispersion appears due to the accumulation of QWS's near
the k-dependent valence band edge of Si. This is an indirect
manifestation of the heavily-doped n-type character of the Si(111)
substrates used here\cite{Speer2006}. The narrow lineshapes of the
energy distribution curves (EDC's; Fig. 2b) and momentum
distribution curves (MDC's; Fig. 2c), and the observations of the
electronic fringes reflect the uniformity of the Ag buffers and
the high resolution of the experiment.

Having discussed the spin-split BiAg surface states and the
quantum-well states in the Ag buffer (without BiAg surface alloy),
we consider their interaction in a BiAg/Ag/Si trilayer, focusing
first on a 17 ML thick Ag buffer (Fig. 2d; i.e. the sample of Fig.
2a covered by the BiAg alloy). The Ag Shockley surface state
disappears and the resulting surface electronic structure agrees
in general with that of the system without Si substrate
(BiAg/Ag(111); no QWS's) but shows intensity modulations in both
the sp$_z$ and p$_x$p$_y$ bands. The energy distribution curves,
extracted from the raw data, clearly evidence band gaps (Fig. 2e).
\begin{figure}
  \centering
  \includegraphics[width = 8 cm]{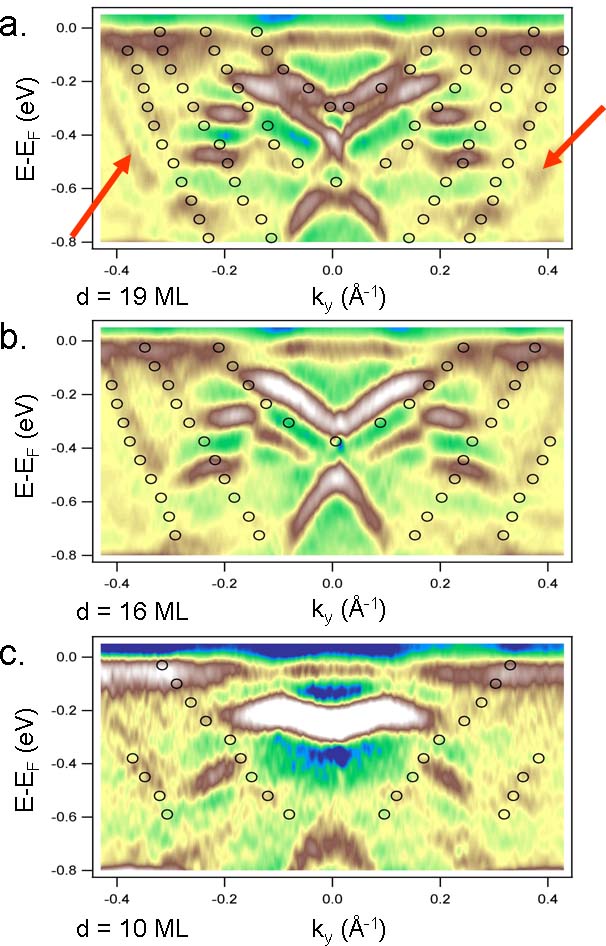}
  \caption{(color online) a, b, and c, second derivative of the ARPES intensity along
  $\bar \Gamma \bar M'$ for three alloy-covered samples at RT with different Ag film thicknesses,
  respectively 19, 16 and 10ML.
  Circles correspond to MDC fits of the QWS observed on the bare Ag thin films of the corresponding thicknesses shifted by
  50-150meV upwards in order to match the remaining parts of the QWS at large k values after Bi deposition (e.g. red arrows).
  White and brown (green) indicate the maxima (minima) of photoemission intensity.}
  \label{fig3}
\end{figure}
The remaining signature of the Ag QWS's (at large k values) and
the gaps in BiAg surface states are clearly seen even at RT in the
second derivative of the ARPES intensities ($d^2 I(E,\vec{
k_\sslash})/dE^2$) for samples with selected Ag film thicknesses
(d = 19, 16, and 10ML) in Fig. 3. The parabolic in-plane
dispersions of the QWS's (circles in Fig. 3) is obtained from
MDC's of Ag/Si(111) with the corresponding Ag thicknesses (as
presented in Fig. 2c). Agreement between the parabolic fits
(uncovered Ag buffer) and the QWS's of the alloyed sample is
obtained after shifting rigidly the parabola by $50-150$meV to
lower binding energies. These shifts can be attributed to the
different reflection properties of the bare Ag surface and of the
BiAg surface alloy or to a possible reorganization of the thin
film upon Bi deposition. The effective masses of the QWS's may
also change. Yet, these fits are to be considered as guides to the
eye. Band gaps are found at the intersection of the QWS fits with
both branches of the surface-alloy bands regardless of their
symmetry or spin, providing strong evidence of their
hybridization. The hybridization is
spin-selective\cite{Didiot2006,Barke2006} thus we can consider in
a first approximation that the QWS are spin-degenerate or their
spin-splitting is small. For thinner Ag buffers (10ML; Fig. 3c),
the number of QWS's is reduced. As a result, the number of band
gaps is also decreased but their widths are larger, in particular
for the p$_x$p$_y$ states.
\begin{figure}
  \centering
  \includegraphics[width = 8.5 cm]{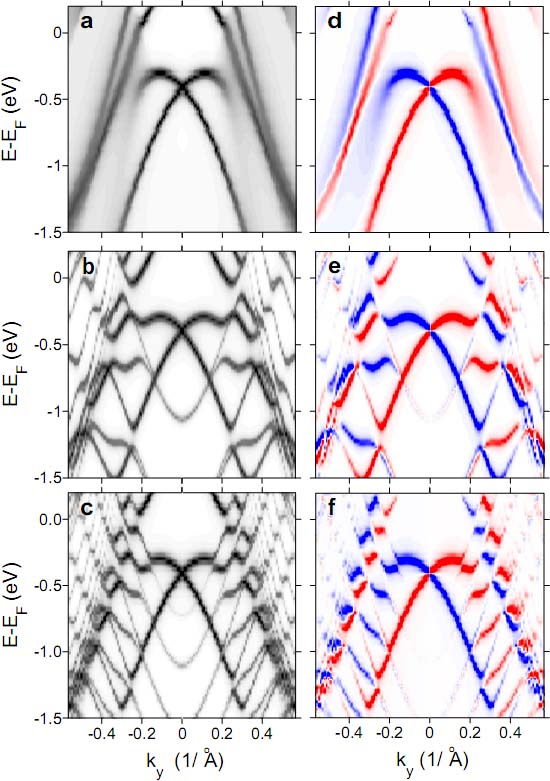}
  \caption{Effect of quantum-well states on the spin-split electronic structure of the Bi/Ag surface alloy,
  as obtained from first-principles electronic-structure calculations.
  a-c: The spectral density at the Bi site is displayed as gray scale
  (with white indicating vanishing spectral weight) for BiAg/Ag(111)
  (a) and BiAg/Ag/Si(111) for Ag buffer thicknesses d = 10 (b) and d = 19 (c).
  The wavevector is chosen as in the experiment (Fig. 3).
  d-f: The spin polarization of the electronic states is visualized by  $\Delta N(E,\vec {k_\sslash})$,
  i.e. the difference of the spin-up and the spin-down spectral density.
  Blue and red indicate positive and negative values, respectively, whereas white is for zero $\Delta N$. }
  \label{fig4}
\end{figure}

To further corroborate the above explanation of the band gaps,
first-principles electronic structure calculations for
BiAg/Ag(111) reported in\cite{Ast2007,Ast2008} were extended.
Since the Ag/Si(111) interface is incommensurate\cite{Speer2006},
we are forced to approximate the Si substrate. Therefore, the
confinement of the Ag QWS's by the Si(111) substrate is mimicked
by replacing Ag bulk layers by repulsive potentials. The latter
provide the complete reflection of the Ag states at the Ag/Si(111)
interface. Note that by this means details of the Ag/Si interface
are roughly approximated and the binding energies of the
theoretical quantum well states may differ from their experimental
counterparts. However, the essential features are fully captured,
as will be clear from the agreement of experiment and theory
discussed below. The systems investigated comprise the BiAg
surface alloy, Ag layers, and the substrate built from hard
spheres (HS; i. e. BiAg/Ag$_{d-1}$/HS(111)). The theoretical
analysis focuses on the wavevector- and spin-resolved spectral
density $N(E,\vec {k_\sslash};\sigma)$ at a Bi site ($\sigma=
\uparrow$or$\downarrow$ is the spin quantum number).
Spin-dependent band gaps are conveniently visualized by displaying
$\Delta N(E,\vec{k_\sslash})=N(E,\vec
{k_\sslash};\uparrow)-N(E,\vec {k_\sslash};\downarrow)$.

For BiAg/Ag(111), the Bi surface states hybridize with Ag bulk
states, resulting in a rather blurred spectral density (Fig. 4a;
compare Fig. 1c for the experiment). For the systems with Si
substrate, focusing here on exemplary results for d = 10 (Fig. 4b)
and 19 (Fig. 4c), quantum well states show up as parabolas
centered at $\bar \Gamma$. The most striking difference to
BiAg/Ag(111) are, however, spin-dependent band gaps at $(E,\vec
{k_\sslash})$ points at which the QWS's would cross the Bi bands.
With increasing thickness of the Ag buffer, the number of gaps (or
QWS's) increases and the width of the gaps decreases. The spectral
densities of the Bi states are slightly less blurred than for
BiAg/Ag(111) because hybridization with Ag states occurs only at
the band gaps, due to quantization. Eventually, we find a shift of
the QWSs' energies upon covering the Ag buffer with the BiAg
alloy, in agreement with experiment. In summary, the
electronic-structure calculations corroborate fully the
experimental findings.

By contrast with what has been observed in Bi thin layers on
silicon\cite{Hirahara2006,Hirahara2007}, Figs. 4e and f clearly
show that the Ag quantum well states are spin-polarized due to the
Rashba effect. Close to $\bar \Gamma$, the branches of opposite
spins of the QWS follow a parabolic dispersion and their momentum
separation decreases with the Ag thickness. This feature is
evident in figures that show spin polarization of the electron
states,  . We now address in particular the electronic structure
at the Fermi level. For d=10ML (Fig. 4e), highly spin-polarized
states show up at $k_\sslash=k_F=0.22$\AA$^{-1}$, with a spin
polarization of about 33\%. On the contrary, a complete gap
appears for d=19 ML (Fig. 4f). These findings imply that the
spin-dependent electronic structure at the Fermi level - and thus
the transport properties - can be drastically modified by the Ag
film thickness.

Our findings for BiAg/Ag/Si(111) trilayers suggest that it is
indeed possible to match systems with large spin-orbit splitting
(here: BiAg/Ag(111)) with a semiconductor substrate. Furthermore,
interfacial properties can be custom-tailored, in the present case
by a single parameter, namely the Ag buffer layer thickness. Within this
respect, multilayer systems which comprise semiconducting Si
layers and Rashba-split subsystems (like BiAg/Ag) may be very
useful in the development of new spintronics devices. Tuning the
band-gap structure at the Fermi level could also be achieved by
chemical means, as was demonstrated for Bi$_x$Pb$_{1-x}$Ag$_2$
mixed alloys grown on Ag(111)\cite{Ast2008}. Peculiar transport
properties and spin Hall effects can be anticipated based on this interface, namely in
nanostructured systems or (Bi-Ag-Si) superlattices.

\begin{acknowledgments}
E.F. acknowledges the Alexander S. Onassis Public Benefit
Foundation for the award of a scholarship. This research was
supported in part by the Swiss NSF and the NCCR MaNEP.
\end{acknowledgments}


\end{document}